\documentclass[english,aps,prl,twocolumn,showpacs,superscriptaddress,groupedaddress,footinbib,reprint,noshowpacs]{revtex4}
\usepackage{graphicx}
\usepackage{amsmath}
\usepackage{comment}

\usepackage{subfigure}

\usepackage{subfigure}
\usepackage{color}
\usepackage{soul}

\begin{document}

\title{Switching the structural force in ionic liquid--solvent mixtures by varying composition}

\author{Alexander M. Smith}
\email{Alexander.Smith@unige.ch} 
\affiliation{Department of Chemistry, Physical and Theoretical Chemistry Laboratory, University of Oxford, Oxford OX1 3QZ, U.K.}
\affiliation{Department of Inorganic and Analytical Chemistry, University of Geneva, 1205 Geneva, Switzerland} 

\author{Alpha A. Lee}
\email{alphalee@g.harvard.edu} 
\affiliation{School of Engineering and Applied Sciences, Harvard University, Cambridge, MA 02138, USA} 

\author{Susan Perkin}
\email{susan.perkin@chem.ox.ac.uk} 
\affiliation{Department of Chemistry, Physical and Theoretical Chemistry Laboratory, University of Oxford, Oxford OX1 3QZ, U.K.}

\begin{abstract}
The structure and interactions in electrolytes at high concentration have implications from energy storage to biomolecular interactions. However many experimental observations are yet to be explained in these mixtures, which are far beyond the regime of validity of mean-field models. Here, we study the structural forces in a mixture of ionic liquid and solvent that is miscible in all proportions at room temperature. Using the surface force balance to measure the force between macroscopic smooth surfaces across the liquid mixtures, we uncover an abrupt increase in the wavelength above a threshold ion concentration. Below the threshold concentration the wavelength is determined by the size of the solvent molecule, whereas above the threshold it is the diameter of a cation--anion pair that determines the wavelength. 
\end{abstract}

\makeatother
\maketitle

Room temperature ionic liquids are revolutionising many application areas, from electrochemical energy storage to lubrication \cite{fedorov2014ionic}. Their wide electrochemical window, chemical stability and low volatility make them ideal electrolytes for electrochemical appliances \cite{hallett2011room,fedorov2014ionic}. Pure ionic liquids are usually mixed with polar organic solvents to lower their viscosity in order to enhance conductivity \cite{mcewen1999electrochemical,zhu2011carbon,yang2013liquid}. However, the effect of solvent concentration on the interfacial structure of ionic liquid-solvent mixtures at high concentration is not yet clear. The combination of ionic liquid and polar solvent is, in some respects, similar to simple inorganic salts in water; e.g. the long range screening observed at very low salt concentrations and recently also demonstrated at high concentrations is similar for ionic liquids and NaCl in water \cite{smith2016electrostatic}. However in other respects ionic liquid electrolytes are distinctly different to standard inorganic electrolytes: (i) Ionic liquids can be mixed with solvents over a far greater range of mole fraction, often fully miscible from 0 - 100\% salt, allowing different regimes to be accessed; (ii) The complex orientation-dependent interactions in ionic liquids lead to non-trivial bulk structure \cite{hayes2015structure} and potentially different electrochemical and physical properties. 

Previous studies have shown that, in many cases, pure ionic liquids form a layered structure near a charged surface with an alternating arrangement of counter-ions and co-ions \cite{perkin2010layering,Merzger2008} and the layer thickness is determined by the dimension of an ion pair. This layering effect gives rise to an oscillatory structural force --- the force between charged surfaces in an ionic liquid oscillates between attraction and repulsion as a function of surface separation \cite{horn1988}. The wavelength of the structural force, determined by the width of those ion layers, sets the location of energy minima between colloidal particles and between macroscopic bodies in ionic liquid solutions. As such it is important for phenomena such as electrowetting \cite{millefiorini2006electrowetting,restolho2009electrowetting,li2013dynamic}, electrokinetic flows \cite{jiang2011electrokinetic,storey2012effects}, and friction/adhesion \cite{perkin2010layering,werzer2012ionic,smith2013quantized,espinosa2013microslips,smith2014molecular}; in each of these examples the microscopic length scale of fluid ``layering'' is critical.  Although controlling the screening length is a common theme in soft matter \cite{evans1999colloidal}, to our knowledge the tuning of the \emph{layer thickness} and the \emph{wavelength} of the structural force has rarely been reported; one instance is the switch in wavelength with the transition from monolayer to bilayer structures in ionic liquids \cite{smith2013monolayerJPCL,cheng2016effect}. 

Here we study the structural force in an ionic liquid mixed with a polar solvent over the full range of mole fractions, from 0\%-100\% added salt, and demonstrate a discontinuous switch in the wavelength of the oscillatory structural force at an intermediate concentration. At low concentrations (below 30 Mol\% salt) the wavelength is constant and determined by the solvent diameter, whereas at high concentrations the wavelength is again constant but determined by the ion pair dimension. This transition is sharp within our experimental resolution. Our measurements also reveal that there are two decay lengths in concentrated electrolytes --- the decay length of the oscillatory structural force and decay length of the long range monotonic screening \cite{smith2016electrostatic}; the two lengths can differ by an order of magnitude. 

\begin{figure}
\includegraphics[scale =0.8]{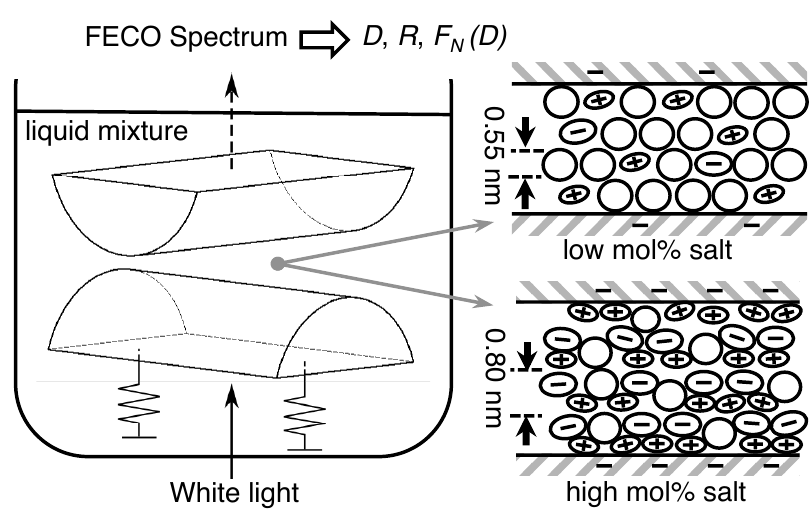}
\caption{Drawing of the experimental geometry in the SFB: mica sheets mounted in crossed-cylinder configuration are brought together in a mixture of ionic liquid and neutral solvent. The distance between the sheets, $D$, their curvature $R$, and their interaction force, $F_N$, are all determined using the spectral fringes of equal chromatic order (FECO). The resulting $F_N$ vs. $D$ profiles reveal the sequential squeeze-out of layers of molecules; cartoons on the right show proffered molecular arrangements at low and at high salt concentration giving rise to solvation force wavelengths determined by solvent and by ion pair respectively.}
\label{drawing}
\end{figure}

Our measurements were carried out using the surface force balance (SFB). The technique and detailed experimental procedures have been described elsewhere \cite{perkin2006forces}, and are shown schematically in Figure  \ref{drawing}. Two back-silvered mica pieces of uniform thickness (1-3 $\mu m$ in these experiments) are glued to cylindrical silica lenses with the mica facing upwards. For reasons of solubility, EPON 1004 (Shell Chemicals) is used to glue mica pieces for ionic liquid-rich solutions, while glucose (Sigma-Aldrich, 99.5\%) is used for the propylene carbonate-rich solutions. The lenses are mounted in the apparatus facing each other in a crossed-cylinder configuration. White-light multiple beam interferometry is used to determine the surface separation $D$ by means of constructive interference fringes of equal chromatic order (FECO) \cite{israelachvili1973thin}. The deflection of a horizontal leaf spring is measured directly via interferometry to determine the surface force $F_N$, with resolution better than $10^{-7} N$. Forces are normalised by the mean local radius of curvature, $R$, of the glued mica surfaces ($R \approx 1 \; \mathrm{cm}$), inferred from the interference pattern, and allows comparison between different contact regions. The measured force $F_N$ is related to the theoretical interaction energy, $E$, between flat surfaces at the same surface separation via the Derjaguin approximation $F_N/R = 2\pi E$ \cite{israelachvili2011intermolecular}. 

The ionic liquid used, 1-butyl-1-methylpyrrolidinium bis[(trifluoromethane)sulfonyl]imide, $\mathrm{[C_4C_1Pyrr][NTf_2]}$ (Iolitec, 99\%), was dried in vacuo ($10^{-2}$ $\mathrm{mbar}$, 70 $^{o}\mathrm{C}$) overnight before preparing the electrolyte solutions. Propylene carbonate (Sigma Aldrich, anhydrous, 99.7\%) was used as received from freshly opened bottles.  

\begin{figure*}
\includegraphics[scale =1]{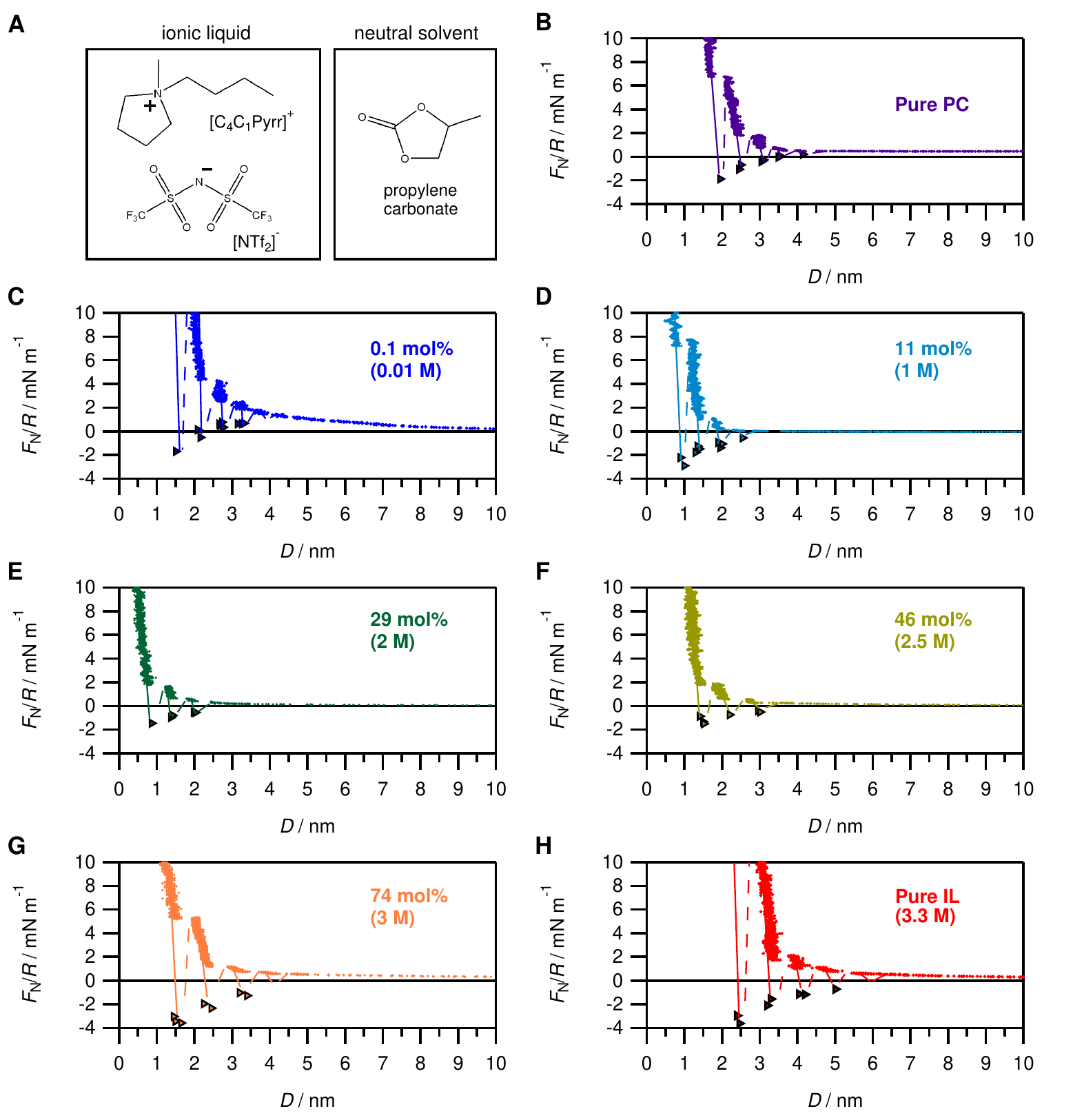}
\caption{(A) Chemical structures and (B-H) measured forces, $F_N$, between two mica surfaces (normalized by radius of curvature, $R$) as a function of surface separation, $D$, across liquids with concentrations ranging from pure propylene carbonate to pure ionic liquid. Concentration values correspond to actual concentrations injected into the SFB rather than derived from a fit to data. Coloured data indicate repulsive forces measured on approach of the surfaces, and triangles indicate points measured from the jump-apart of the surfaces from adhesive minima. Lines are a guide to the eye to show the oscillatory nature of the forces, with solid lines through the measured regions and dashed lines through regions inaccessible to measurement.}
\label{full_exp_data}
\end{figure*}

Figure \ref{full_exp_data} shows the measured force between atomically smooth mica sheets for concentrations ranging from pure propylene carbonate (0\% ionic liquid; Fig.1B) up to 100\% ionic liquid (Fig.1H). Interaction forces in this system can involve a weak monotonic component extending beyond the layering region. We have discussed this aspect elsewhere \cite{smith2016electrostatic}; here we focus on the layering region where the force minima, and thus layer width, can be resolved.

We first examine the behaviour of pure propylene carbonate liquid, as shown in Fig. \ref{full_exp_data}B. The oscillatory force measured, corresponding to layering, is in agreement with previous surface force measurements of non-aqueous polar \cite{christenson1983direct,christenson1984dlvo,christenson1985forces} and non-polar \cite{horn1980direct,horn1981direct} liquids confined between mica surfaces. The wavelength is $\sim0.55  \; \mathrm{nm}$, similar to the molecular size of propylene carbonate. The weak exponential tail is due to dissociation of potassium ions from the mica surface. Fitting to DLVO theory yields a measured Debye length of $112 \mathrm{nm}$, which corresponds to a 1:1 electrolyte concentration of $6 \times 10^{-6} M$ and consistent with dissociated potassium ions and ions arising from dissolved gas molecules. 

Addition of $10^{-2} \mathrm{M}$ of $\mathrm{[C_4C_1Pyrr][NTf_2]}$ to the propylene carbonate, as shown in Fig. \ref{full_exp_data}C, results in a qualitatively similar force profile; wavelength corresponding to pure propylene carbonate solvent, $0.55 \mathrm{nm}$, is observed. These oscillatory structural forces appear to be additive to the double layer force, which has a decay length of 3 $\mathrm{nm}$ ($0.8 \times 10^{-2} \mathrm{M}$ assuming a 1:1 electrolyte) agreeing with the nominal solution concentration. Similar behaviour continues up to a concentration of $\mathrm{2 M}$ (Fig. 1D-E), albeit with a weaker magnitude.

The results for the $\mathrm{2.5 M}$ solution are shown in Fig. \ref{full_exp_data}F. Strikingly, the wavelength changes abruptly at this concentration: the wavelength is $\sim0.8 \mathrm{nm}$. Increasing the concentration further to $3 \mathrm{M}$ (Fig. \ref{full_exp_data}G) results in a wavelength of $\sim0.8 \mathrm{nm}$ again, with the force oscillations now more pronounced. Finally, we studied the pure ionic liquid, with no propylene carbonate solvent, giving rise to an intrinsic ion concentration of $3.3 \mathrm{M}$; this is shown in Fig. \ref{full_exp_data}H.  The distance between force minima is determined as $0.8 \pm 0.04 \mathrm{nm}$, which agrees with previous SFB \cite{smith2013monolayer,smith2013monolayerJPCL} and atomic force microscope \cite{hayes2009pronounced} measurements for this ionic liquid. 

\begin{figure}
\includegraphics[scale = 0.8]{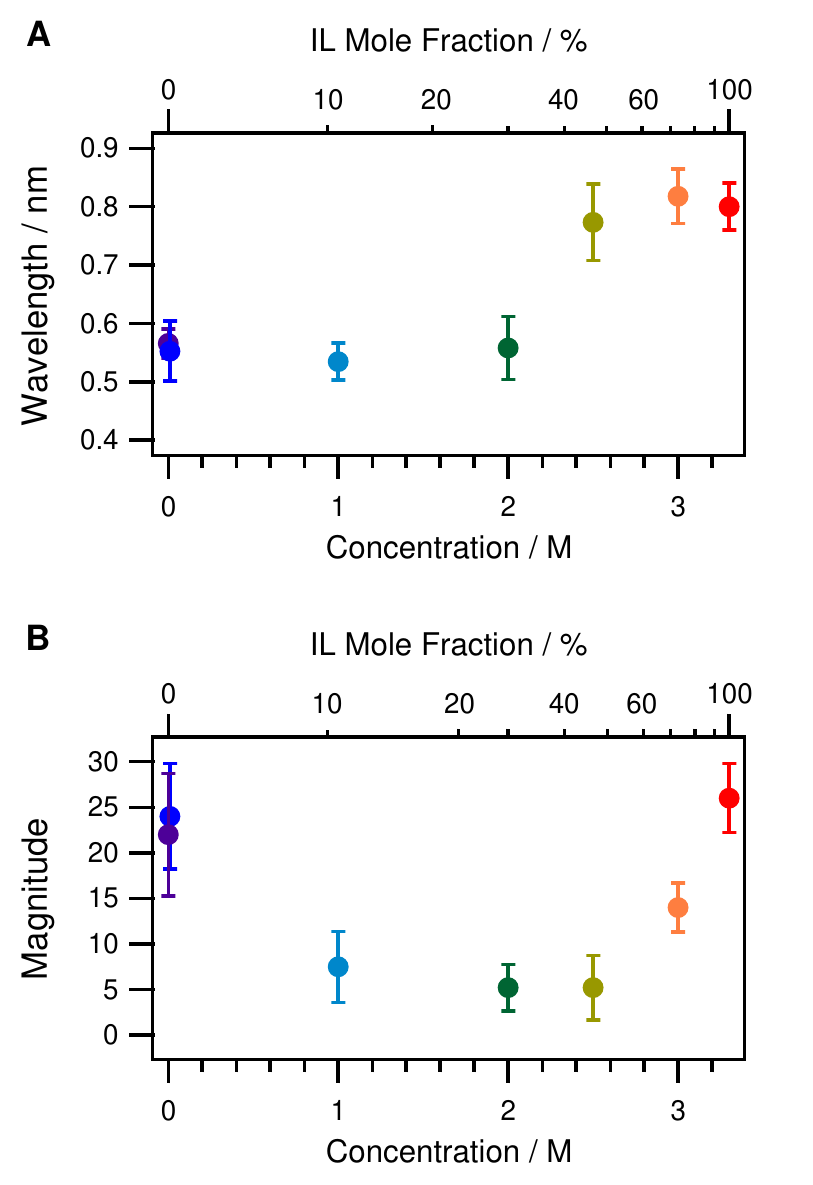}
\caption{(A) Wavelength of the measured oscillatory forces, corresponding to each of the concentrations shown in Fig. \ref{full_exp_data} B-H. Values were calculated using the measured distances between adjacent minima and between adjacent maxima in the force profiles. (B) Magnitude of the oscillatory force at each of the concentrations studied. Values were calculated from the pre-exponential factor associated with the decay of adhesive minima in the force oscillations for each concentration.}
\label{sum}
\end{figure}

Figure \ref{sum} summarises our key results: the wavelength of the structural force changes from the size of a solvent molecule in a dilute ionic liquid solution to the size of a cation-anion pair in a concentrated solution. To within experimental error, this transition is sharp. The magnitude of the force reaches a minimum at the threshold concentration associated with this transition. This force magnitude minimum could be heuristically interpreted as the presence of ions (solvent) ``disrupting'' the structure of the solvent (ions). 

\begin{figure}
\includegraphics[scale = 0.8]{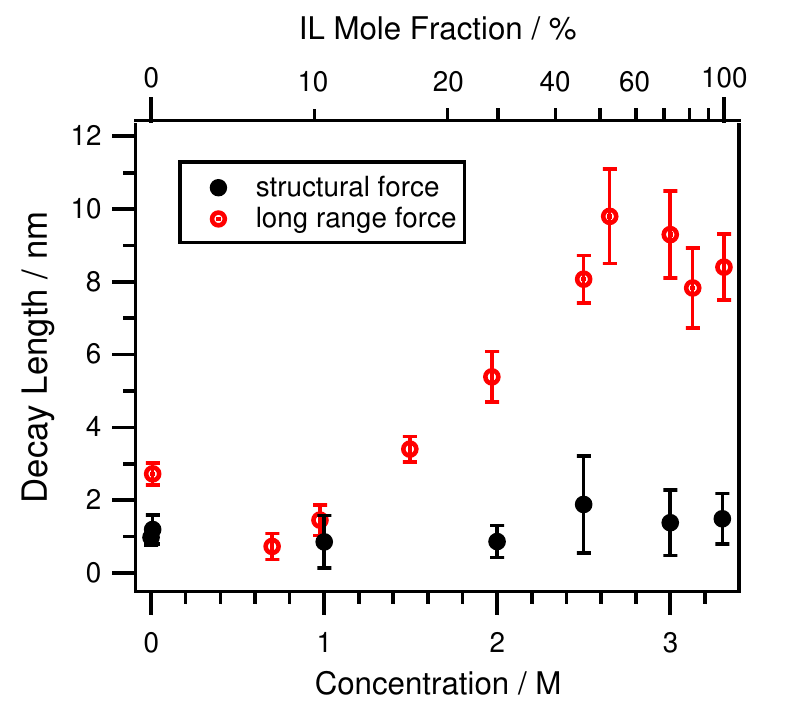}
\caption{Two decay lengths are required to characterise the interaction between charged bodies in concentrated electrolytes. The decay length of the oscillatory structural force is determined from an exponential fit to the minima of oscillations shown in Fig. \ref{full_exp_data} B-H. The decay of the long-range electrostatic force is determined from an exponential fit to the long range tail. This was recently show to demonstrate an anomalous increase at high concentration \cite{smith2016electrostatic}. The long-range screening length measured in pure PC, not shown, is $112 \mathrm{nm}$. } 
\label{fig3}
\end{figure}

Another aspect of this oscillatory structural force is the characteristic decay length of the envelope enclosing the oscillatory force. The envelope can be fitted to an exponential decay and Figure \ref{fig3} shows that the decay length is approximately constant ($\sim 0.9 \; \mathrm{nm}$) up to $2 \mathrm{M}$, then increases to $\sim 1.5 \; \mathrm{nm}$ beyond $2 \mathrm{M}$, and remains the same up to pure ionic liquids. This transition in the decay length of the oscillatory envelope occurs at the same concentration as the jump in oscillation wavelength from the dimension of a solvent molecule to the dimension of an ion pair. The apparent correlation between the oscillation wavelength and decay length of the structural force suggests that the oscillatory structural force can be described by a single lengthscale which determines both the wavelength and decay. We note that the decay length of this oscillatory structural force is distinctly different to the decay length of the long-ranged electrostatic screening reported by us previously for the same liquids \cite{smith2016electrostatic} (Figure \ref{fig3}). We showed previously that the electrostatic screening length, $\lambda_s$, estimated by fitting the surface forces beyond the oscillatory region to an exponential decay, can be much longer than the Debye-Huckel screening length computed using the nominal ion concentration and demonstrates an anomalous increase with concentration for concentrated electrolytes \cite{smith2016electrostatic}. In the pure ionic liquid the structural force is short-ranged whereas the electrostatic screening length is an order of magnitude larger than the ion diameter.

The existence of two decay lengths, and the fact that the decay length of the short-ranged, oscillatory component is independent of the electrolyte concentration except the abrupt jump whereas the decay length of the long range force continuously varies with concentration (Figure \ref{fig3}), suggest that the oscillatory structural force may have origin in steric packing of a strongly confined system rather than long ranged electrostatic interactions. Focusing on steric effects, in a mixture of ``big'' and ``small'' hard-sphere particles, an abrupt crossover in the oscillatory wavelength as a function of concentration is predicted \cite{grodon2004decay} and observed in experiments with colloidal particles \cite{baumgartl2007experimental,statt2016direct}; the ion pair diameter is almost twice the solvent diameter, well in the regime where this structural transition is expected to occur \cite{grodon2004decay}.  This abrupt transition is not apparent in earlier experiments with big and small non-polar molecules \cite{vanderlick1991forces}, possibly because the range of compositions investigated was narrower and the SFB experiments had lower resolution. 

We note that in X-ray reflectivity experiments \cite{mezger2015solid}, the structure of ionic liquid adjacent to a single charged interface has been studied as a function of dilution and confirms an oscillatory charge density decay at higher concentrations (corresponding to the Kirkwood line \cite{attard1993asymptotic,leote1994decay}). However, the solvent-determined periodicity in our SFB ``squeeze-out'' experiments is not apparent in the analysis of those experiments. This difference is likely due to the fact the analysis \cite{mezger2015solid} has not accounted for the possibility of two decay lengths, one for the correlation between ``bigger'' ion pairs and one between ''smaller'' solvent molecules  \cite{statt2016direct}. 

Another possible, though less likely, mechanism that may give rise to a switch in the structural force is an adsorption mechanism. In systems where a bulk electrolyte reservoir is in equilibrium with a confined system, it is well known that the concentration of ionic species in the confined system is different to the bulk electrolyte concentration (the Donnan effect, e.g. \cite{dahirel2009ion}). Future work using experimental setups that can track the composition of the fluid in the mica-mica slit \cite{kienle2016density} will shed light on whether the composition of the electrolyte changes abruptly as a function of bulk concentration.  

In summary, we studied the force between mica sheets across mixtures of neutral propylene carbonate and an ionic liquid salt at concentrations ranging from pure propylene carbonate up to pure ionic liquid. The  short-range structural force is oscillatory and indicative of layering, and the wavelength of the force oscillations changes abruptly as the ion concentration is increased. Moreover, the decay of the oscillatory envelope is connected to the oscillation wavelength, but entirely different (with different concentration dependence) to the long-range screening reported recently in the same liquid mixtures. Although results are reported for the $\mathrm{[C_4C_1Pyrr][NTf_2]}$--propylene carbonate system, we believe this phenomenology is general for mixtures of ionic liquid with non-aqueous solvents. Previous SFB experiments have shown that the oscillation wavelength of the force profile for pure ionic liquids is approximately an ion pair diameter regardless of ion chemistry as long as the length of the alkyl chains is below the threshold for forming a bilayer structure \cite{perkin2011self}. Therefore the choice of this particular ionic liquid is completely arbitrary--- $\mathrm{[C_4C_1Pyrr][NTf_2]}$ is chosen only because it is one of the ionic liquids that is completely miscible with propylene carbonate at all compositions at room temperature.

Our results point towards a new avenue to tune interfacial phenomena such as adhesion, lubrication, wetting and electrokinetic flows with ionic liquid solutions. In particular, as the friction response of ionic liquid-lubricated systems display distinct friction-load regimes depending on the number of confined ion layers \cite{smith2013quantized}, tuning the lengthscale of layering is one significant step towards addressable lubricant design. Moreover, we hope our approach will open up the horizon of rational electrolyte design via engineering the wavelength and decay of short-range structural forces, perhaps independently from the long-range screening length. In particular, multiple Òswitch settingsÓ for oscillation wavelength may be possible if more length scales are introduced by using a mixture of solvents and salts. 

\acknowledgments
The authors are very grateful to R Evans for many insightful discussions. AMS is supported by a Doctoral Prize from the EPSRC. AAL is supported by a UK-US Fulbright Fellowship to Harvard University and the George F. Carrier Fellowship. SP is supported by The Leverhulme Trust (RPG-2015-328) and the ERC (under Starting Grant LIQUISWITCH). 

\bibliography{IL_PC_ref} 
\end{document}